\documentclass[reprint, amsmath, amssymb, amsfonts, aps, a4paper, dvipdfmx]{revtex4-1}

\usepackage[utf8]{inputenc}
\usepackage{graphicx}
\usepackage{bxtexlogo}
\usepackage{type1cm}
\usepackage{diagbox} 
\usepackage{mathtools}
\usepackage[toc,page]{appendix}
\usepackage{color}
\usepackage{ulem}

\newcommand{\dfracp}[2]{\dfrac{\partial #1}{\partial #2}}

\begin{document}

\title{Inferring coupling strength and natural frequency distribution in coupled Stuart-Landau oscillators using linear response}
\author{Moonseok Choi}
\email{choi.m.ab@m.titech.ac.jp}
\affiliation{Department of Systems and Control Engineering, Institute of Science Tokyo, Tokyo 152-8552, Japan}
\author{Yoshiyuki Y. Yamaguchi}
\email{yyama@amp.i.kyoto-u.ac.jp}
\affiliation{Graduate School of Informatics, Kyoto University, Kyoto 606-8501, Japan}
\begin{abstract}
  We propose a framework to infer the coupling strength and the natural frequency distribution
  in a coupled Stuart-Landau oscillator system with a large population.
  The inference method uses observation of linear response
  of a macroscopic quantity and of an oscillator.
  We first solve the direct problem on the response
    with transforming the system into the phase-amplitude equations.
  Solving the inverse problem, we show that the coupling strength is inferred
  from observation of an oscillator and the natural frequency distribution from macroscopic responses.
  The proposed method requires only one-dimensional observation in the two-dimensional Stuart-Landau system. Validity of the inference theory is examined by numerical simulations.
\end{abstract}

\maketitle


\section{Introduction}
\label{sec:Introduction}

Collective rhythmic phenomena are ubiquitously observed in nature including
flashing of fireflies \cite{Buck1968} and flog choruses \cite{Aihara2014}.
They are modeled by coupled
limit-cycle oscillators, such as the Stuart-Landau (SL) equation \cite{LandauCollected, Stuart1960, Kuramoto1984}, van der Pol oscillator \cite{vanderPol1926}, FitzHugh-Nagumo oscillator \cite{FitzHugh1961, Nagumo1962},
and the Hodgkin-Huxley model \cite{Hodgkin1952}.
The Stuart-Landau oscillator often arises as the normal form of Hopf bifurcations in a wide range of systems, including chemical
oscillators such as the Belousov-Zhabotinsky reaction
\cite{Ipsen1997}. The van der Pol oscillator, initially developed to
describe electrical circuits, has become a widely used model in
biological contexts \cite{Pikovsky2001}, while the FitzHugh-Nagumo
model provides a simplified description of neuronal
excitability. Furthermore, the Hodgkin-Huxley model \cite{Hodgkin1952}
offers a detailed framework for understanding action potential
generation in neurons, which is a vital component in neuroscience.
  
A coupled limit-cycle oscillator model
is reduced to a coupled phase-oscillator model
by focusing on dynamics around the limit cycle
and using phase reduction techniques \cite{Kuramoto1984, Winfree2001, Nakao2016}.
The phase response curve (or the phase resetting curve)
has been shown to be critical for modeling synchronization in real neurons
\cite{Galan2005, Smeal2010}. 
A simple and paradigmatic coupled phase-oscillator model
is the Kuramoto model \cite{Kuramoto1975}
and there are several generalizations (see \cite{Acebron2005} for instance). 

In studies of coupled phase-oscillator systems,
there is a high demand to specify the underlying dynamical system
responsible for observed rhythmic phenomena in the real world.
There are works which have focused on reconstructing coupling functions
from microscopic multivariate time-series data
\cite{Miyazaki2006, Tokuda2007, Penny2009, Kralemann2011, Tokuda2019, Arai2021}.

However, in a large system effectively approaching infinity,
observing time series of all oscillators is hard.
This obstacle induces another issue that
the natural frequency distribution is left unknown,
while it changes bifurcation type, for instance, a critical exponent
\cite{Kuramoto1975,Crawford1995,Strogatz2000,Chiba2011,Komarov2013,Komarov2014,Chiba2015,Daido2015}
and anomalous transitions induced by asymmetry of distributions
\cite{Terada2017,Yoneda2020}.
In such a system, it is crucial to have an inference method
which observes only macroscopic quantities.
A linear response theory \cite{Terada2020} in the equation of continuity,
which describes dynamics in the large population limit,
enables us to infer the coupling function and the natural frequency distribution
from observation of macroscopic responses \cite{Yamaguchi2024}.

The inference method using macroscopic responses is useful in large systems,
but the previous method has been developed in coupled phase-oscillator systems \cite{Yamaguchi2024}.
  In other words, it assumes that a limit-cycle oscillator stays on its original limit-cycle, although couplings and an external force kick in.
We aim to address this gap by including deviations from the limit-cycles.
Our main goal is to propose a new idea to infer a coupled limit-cycle oscillator system
from linear responses,
and we focus on tractable SL systems as a first testbed.
In the SL systems the polar coordinates perform the phase-amplitude reduction,
which we use in our inference method,
but the phase-amplitude reduction framework has been proposed to incorporate amplitude deviations in addition to phase dynamics \cite{Wilson2016, Shirasaka2017, Mauroy2018, Wilson2020, Mircheski2023} in general limit-cycle systems.




We clarify assumptions introduced in the considering coupled SL system.
(i) The natural frequencies are close to each other.
(ii) The number of oscillators is sufficiently large, effectively approaching infinity.
(iii) The couplings are all-to-all with a small positive coupling constant
so as to have nonsynchronized state.
(iv) The coupling and the external force are applied to only one variable included in a SL oscillator.
The assumptions (i) and (ii) are basic to perform the phase reduction
and to construct the equation of continuity.
The assumption (iii) is used in inference of coupled phase-oscillator systems
\cite{Tokuda2007, Yamaguchi2024}.
The assumption (iv) is inspired by the FitzHugh-Nagumo model,
whose variables include the membrane voltage.
We also introduce a restriction on observation:
we observe only the forced variable assumed in (iv).
This restriction enriches applicability of our theory.
We infer the unknown quantities, the coupling constant and the natural frequency distribution,
from the restricted observation.
The coupling constant is estimated by analyzing the deviation of
an oscillator from the limit cycle,
which reflect the system's amplitude response under external forcing.
To infer the natural frequency distribution,
we apply the linear response theory \cite{Terada2020, Yamaguchi2024, Daido2015, Sakaguchi1988}
to a macroscopic quantity including phase and amplitude responses.

The remainder of this article is organized as follows.
In Sec.~\ref{sec:Model}, we present the Stuart-Landau model
and derive the phase-amplitude equations.
Section~\ref{sec:DirectProblemResponse} discusses the direct problem
which predicts the linear response of a macroscopic quantity and of an individual oscillator.
The inference formula is provided in Sec.~\ref{sec:InverseProblemInference}
by solving the inverse problem.
Section~\ref{sec:NumericalTests} presents numerical simulations to validate
the proposed inference method.
Finally, Sec.~\ref{sec:Conclusion} summarizes our conclusions and perspectives.

\section{Model}
\label{sec:Model}

We introduce a coupled SL oscillator system
and derive the phase-amplitude equations,
which are used in the inference.

\subsection{Stuart-Landau oscillators}

The general form of the SL equation is given by
\begin{equation}
  \dfrac{dX}{dt}
  = a X - b X |X|,
  \qquad
  X,a,b\in\mathbb{C}.
  \label{eq:generalSL}
\end{equation}
We set $a=\alpha +i\omega ~(\alpha>0,\omega\in\mathbb{R})$,
while we assume $b=1$ for simplicity. 
The complex variable $X$ is rewritten as $X=x+iy$ by using real variables $x$ and $y$, and the resulting equations are
\begin{equation}
  \begin{split}
    & \dot{x} = \alpha x - \omega y - (x^{2}+y^{2}) x, \\
    & \dot{y} = \alpha y + \omega x - (x^{2}+y^{2}) y. \\
    \label{eq:SLmodel-mono}
  \end{split}
\end{equation}
Here, for an arbitrary variable $z$, $\dot{z}$ represents the time derivative as
$\dot{z}=dz/dt$.

Following the assumptions (iii) and (iv) mentioned in Sec.~\ref{sec:Introduction},
we consider the following coupled SL oscillators
\begin{equation}
  \begin{split}
    & \dot{x}_{j}
      = \alpha x_{j} - \omega_{j} y_{j} - (x_{j}^{2}+y_{j}^{2}) x_{j}
      + \dfrac{2K}{N} \sum_{k=1}^{N} x_{k} + P(t), \\
    & \dot{y}_{j}
      = \alpha y_{j} + \omega_{j} x_{j} - (x_{j}^{2}+y_{j}^{2}) y_{j}, 
  \end{split}
  \label{eq:SLmodel}
\end{equation}
where the $j$-th oscillator has the dynamical variables $(x_{j},y_{j}), ~j = 1,\dots, N$ for a population of size \(N\) and the natural frequency $\omega_{j}$.
The natural frequencies $\omega_{j}$ follow the natural frequency distribution $g(\omega)$.
$K$ is the coupling constant assumed to be positive.
The external force $P(t)$ is set as
\begin{equation}
  P(t) = 2 h \cos(\omega_{\rm ex}t).
  \label{eq:ExternalForce}
\end{equation}
In what follows, we assume smallness of the external force,
$1\gg h\gg 1/\sqrt{N}$, to make the response quasi linear.
The lower bound of $h$ ensures distinguishability of response
from finite-size fluctuation  of order $O(1/\sqrt{N})$.
The factor $2$ in the coupling constant and the external force is for convenience
in later calculations.

We suppose that the coupling constant $K$ and the natural frequency distribution $g(\omega)$
are unknown as well as the linear growth ratio $\alpha$.
Our aim is to infer them from observation of time series of one of $x_{j}(t)$
and a macroscopic quantity
\begin{equation}
  \bar{x} = \dfrac{1}{N} \sum_{k=1}^{N} x_{k}.
\end{equation}
Note that the intrinsic value of $\bar{x}$ is zero from the assumption (iii)
and $|\bar{x}|=O(h)$.

\subsection{Phase-amplitude equations}
We rewrite Eq.~\eqref{eq:SLmodel} in the phase-amplitude equations.
They are systematically obtained \cite{Wilson2016, Shirasaka2017}, but in
Eq.~\eqref{eq:SLmodel}, the phase-amplitude equations are simply
derived by taking the polar coordinates
$(x_{j},y_{j})=(r_{j}\cos\theta_{j},r_{j}\sin\theta_{j})$ as
\begin{equation}
  \begin{split}
    & r_{j}\dot{\theta}_{j}
      = r_{j} \omega_{j} - 2K \bar{x} \sin\theta_{j} - P(t) \sin\theta_{j}, \\
    & \dot{r}_{j} = \alpha r_{j} - r_{j}^{3} + 2K \bar{x} \cos\theta_{j} + P(t) \cos\theta_{j}. \\
  \end{split}
\end{equation}
The coupling term and the external force term are of $O(h)$,
and the radius $r_{j}$ can be expanded as $r_{j}=r_{0j}+r_{1j}$, where $|r_{1j}|=O(h)$.
The zeroth order equation for $r_{0}$ is
\begin{equation}
  \dot{r}_{0j} = \alpha r_{0j} - r_{0j}^{3}
  \label{eq:r0}
\end{equation}
and the solution is $r_{0j}(t)\to \sqrt{\alpha}$ in the limit $t\to\infty$.
The equations for $\theta_{j}$ and $r_{1j}$ are
\begin{equation}
  \dot{\theta}_{j}
  = \omega_{j} - \frac{2}{\sqrt{\alpha}} K\bar{x} \sin\theta_{j}
    -  \frac{1}{\sqrt{\alpha}}P(t) \sin\theta_{j} + O(h^{2})
  \label{eq:EOM-theta}
\end{equation}
and
\begin{equation}
  \dot{r}_{1j}
  = -2\alpha r_{1j} + 2K\bar{x}\cos\theta_{j} + P(t)\cos\theta_{j} + O(h^{2}).
  \label{eq:EOM-r1}
\end{equation}
This expansion induces the expansion of $\bar{x}$ as
\begin{equation}
  \bar{x} = \bar{x}_{\theta} + \bar{x}_{r},
  \label{eq:barx}
\end{equation}
where
\begin{equation}
  \bar{x}_{\theta} = \dfrac{1}{N} \sum_{k=1}^{N} \sqrt{\alpha} \cos\theta_{k},
  \qquad
  \bar{x}_{r} = \dfrac{1}{N} \sum_{k=1}^{N} r_{1k}\cos\theta_{k}.
  \label{eq:definition-xtheta-xr}
\end{equation}
In Sec.~\ref{sec:DirectProblemResponse}
we derive theoretically the formula of $\bar{x}_{\theta}$ and $\bar{x}_{r}$
up to $O(h)$.

\subsection{Further simplified phase-amplitude equations}

We assume that the macroscopic quantity $\bar{x}$ oscillates
with the frequency $\omega_{\rm ex}$ of the external force in the limit $t\to\infty$.
Indeed, it is true for coupled phase-oscillator systems \cite{Terada2020}.
This assumption induces two susceptibilities $\chi_{\rm c}(\omega_{\rm ex})$
and $\chi_{\rm s}(\omega_{\rm ex})$ defined by
\begin{equation}
  \bar{x}
  \xrightarrow{t\to\infty}
  h \left[ \chi_{\rm c} \cos(\omega_{\rm ex}t)
    + \chi_{\rm s} \sin(\omega_{\rm ex}t) \right]
  + O(h^{2})
  \label{eq:xt-chi}
\end{equation}
for a sufficiently small $h$. We remark that the prefactor $2$ vanishes
by taking the averaging over time which omits rapidly oscillating terms.
The resulting equation is of a standard form in a coupled phase-oscillator system.
The linear response coefficients $\chi_{\rm c}$ and $\chi_{\rm s}$ are comparable
with the linear response discussed in \cite{Terada2020}.

From the assumption (i) mentioned in Sec.~\ref{sec:Introduction},
the standard deviation of $g(\omega)$ is much smaller than the mean.
In Eqs.~\eqref{eq:EOM-theta} and \eqref{eq:EOM-r1},
varying $\omega_{\rm ex}$ around the mean $\mu$ and applying the averaging method again,
which neglects rapidly oscillating terms, we have respectively
\begin{equation}
  \begin{split}
    \dot{\theta}_{j}
    = \omega_{j} - \frac{h}{\sqrt{\alpha}}[ A \sin(\theta_{j}-\omega_{\rm ex}t)
      + B \cos(\theta_{j}-\omega_{\rm ex}t) ]
    + O(h^{2})
  \end{split}
  \label{eq:EOM-theta-2}
\end{equation}
and
\begin{equation}
  \dot{r}_{1j}
  = -2\alpha r_{1j}
  + h [ A \cos(\theta_{j}-\omega_{\rm ex}t)
  - B \sin(\theta_{j}-\omega_{\rm ex}t) ]
  + O(h^{2}),
  \label{eq:EOM-r-2}
\end{equation}
where
\begin{equation}
  A(\omega_{\rm ex}) = 1 + K \chi_{\rm c}(\omega_{\rm ex}),
  \qquad
  B(\omega_{\rm ex}) = K \chi_{\rm s}(\omega_{\rm ex}).
  \label{eq:A-B}
\end{equation}

\section{Direct problem: Response}
\label{sec:DirectProblemResponse}

The two susceptibilities $\chi_{\rm c}(\omega_{\rm ex})$ and $\chi_{\rm s}(\omega_{\rm ex})$
are obtained from observed times series $\bar{x}(t)$ as
\begin{equation}
  \begin{split}
    & \chi_{\rm c}(\omega_{\rm ex})
      \simeq \dfrac{2}{h(t_{2}-t_{1})} \int_{t_{1}}^{t_{2}} \bar{x}(t) \cos(\omega_{\rm ex}t) dt, \\
    & \chi_{\rm s}(\omega_{\rm ex})
      \simeq \dfrac{2}{h(t_{2}-t_{1})} \int_{t_{1}}^{t_{2}} \bar{x}(t) \sin(\omega_{\rm ex}t) dt, \\
  \end{split}
  \label{eq:chic-chis}
\end{equation}
where $t_{1}$ and $t_{2}~(t_{1}<t_{2})$ are sufficiently large.
We make a connection between the observed susceptibilities, $\chi_{\rm c}$ and $\chi_{\rm s}$,
and the unknown quantities, $K$ and $g(\omega)$ in Sec.~\ref{Susceptibilities}.
The two susceptibilities are not independent however due to the Kramers-Kronig relation
\cite{Kronig1926,Kramers1927}, which gives one from the other.
We need one more information to infer $K$ in addition to $g(\omega)$.
We will fill this gap by observing deviation from the limit cycle for an oscillator discussed in Sec.~\ref{sec:Deviation}.

\subsection{Susceptibilities}
\label{Susceptibilities}
The response $\bar{x}_{\theta}$ can be computed
by solving Eq.~\eqref{eq:EOM-theta-2}
through the equation of continuity in the limit $N\to\infty$.
See Appendix ~\ref{sec:LinearResponsePhase}.
Remembering that the factor $\sqrt{\alpha}$ is canceled out
between the definition of $\bar{x}_{\theta}$ in Eq.~\eqref{eq:definition-xtheta-xr}
and the effective force strength in Eq.~\eqref{eq:EOM-theta-2},
the response is
\begin{equation}
  \begin{split}
    \bar{x}_{\theta}
    \xrightarrow{t\to\infty}
    h \dfrac{\pi}{2}
    & \left[ (Ag - BH[g]) \cos(\omega_{\rm ex}t) \right. \\
    & \left. + (AH[g]+Bg) \sin(\omega_{\rm ex}t) \right] + O(h^{2}).
  \end{split}
  \label{eq:barx-theta}
\end{equation}
Here, $H[g]$ is the Hilbert transform of $g$ defined by
\begin{equation}
  H[g](\omega)
  = \dfrac{1}{\pi} {\rm PV} \int_{-\infty}^{\infty} \dfrac{g(\omega')}{\omega-\omega'} d\omega'.
\end{equation}

The response $\bar{x}_{r}$ is obtained by solving Eq.~\eqref{eq:EOM-r-2} directly
and replacing the $N$-body average with the distribution average.
After straightforwad computations (see Appendix \ref{sec:LinearResponseAmplitude}), we have
\begin{equation}
  \begin{split}
    \bar{x}_{r}
    \xrightarrow{t\to\infty}
    h 
    & \left[ (AJ_{0}-BJ_{1}) \cos(\omega_{\rm ex}t) \right. \\
    & \left. + (AJ_{1}+BJ_{0}) \sin(\omega_{\rm ex}t) \right] + O(h^{2}),
  \end{split}
  \label{eq:barx-r}
\end{equation}
where
\begin{equation}
  \begin{split}
    & J_{0}(\omega_{\rm ex})
      = \int_{-\infty}^{\infty}
        \dfrac{g(\omega)}{4\alpha^{2}+(\omega_{\rm ex}-\omega)^{2}}\, d\omega, \\
    & J_{1}(\omega_{\rm ex})
      = \dfrac{1}{2}
        \int_{-\infty}^{\infty}
        \dfrac{(\omega_{\rm ex}-\omega)g(\omega)}{4\alpha^{2}+(\omega_{\rm ex}-\omega)^{2}}\, d\omega.
  \end{split}
\end{equation}

The two partial responses $\bar{x}_{\theta}$, Eq.~\eqref{eq:barx-theta},
and $\bar{x}_{r}$, Eq.~\eqref{eq:barx-r}, give the total response $\bar{x}$.
Looking back Eqs.~\eqref{eq:xt-chi} and \eqref{eq:A-B},
we have self-consistent equations for $\chi_{\rm c}$ and $\chi_{\rm s}$ as
\begin{equation}
  \begin{split}
    & \chi_{\rm c} = (1+K\chi_{\rm c})\left( \dfrac{\pi}{2} g + J_{0} \right)
    - K\chi_{\rm s} \left( \dfrac{\pi}{2} H[g] + J_{1} \right), \\
    & \chi_{\rm s} = (1+K\chi_{\rm c})\left( \dfrac{\pi}{2} H[g] + J_{1} \right)
    + K\chi_{\rm s} \left( \dfrac{\pi}{2} g + J_{0} \right). \\
  \end{split}
  \label{eq:SC-chic-chis}
\end{equation}
We underline that the unknown quantities $K$ and $g(\omega)$,
as well as the linear growth ratio $\alpha$,
are included in the right-hand side of the above equations.

\subsection{Deviation from the limit cycle}
\label{sec:Deviation}
We select an p-th oscillator to observe the time series $x_{\rm p}(t)$.
We can estimate the frequency $\omega_{\rm p}$
and the linear growth ratio $\alpha$ without external force:
The period of $x_{\rm p}(t)$ must be $2\pi/\omega_{\rm p}$ from Eq.~\eqref{eq:EOM-theta-2}
and the amplitude of $x_{0{\rm p}}(t)$ converges to $\sqrt{\alpha}$ from Eq.~\eqref{eq:r0}
for $h=0$.
Then, we apply an external force, $h>0$, which gives the time series
\begin{equation}
  x_{\rm p}(t) = [\sqrt{\alpha} + r_{1 {\rm p}}(t)] \cos\theta_{\rm p}(t).
  \label{eq:xp-thetap}
\end{equation}
By setting the frequency of the external force as $\omega_{\rm ex}=\omega_{\rm p}$,
the maximum value $r_{1}^{\rm max}$ is expressed by
\begin{equation}
  r_{1}^{\rm max} = \dfrac{h}{2\alpha} \sqrt{A^{2}+B^{2}}
  = \dfrac{h}{2\alpha} \sqrt{(1+K\chi_{\rm c})^{2}+(K\chi_{\rm s})^{2}}
  \label{eq:r1-max}
\end{equation}
as shown in Eq.~\eqref{eq:r-max},
where $\chi_{\rm c}$ and $\chi_{\rm s}$ are the values at $\omega_{\rm ex}=\omega_{\rm p}$.
  From Eq.~\eqref{eq:xp-thetap}, we have $|x_{\rm p}(t)|\leq \sqrt{\alpha}+r_{1{\rm p}}(t)$,
  and the maximum value of $r_{1{\rm p}}(t)$, corresponding to $r_{1}^{\rm max}$,
  is estimated by observing the maximum value of $x_{\rm p}(t)$.

\section{Inverse problem: Inference}
\label{sec:InverseProblemInference}
The inference proceeds in two steps.
First, the coupling strength $K$ is determined
from the amplitude response of a selected oscillator,
which provides an independent relation between $K$
and the measured susceptibilities.
Second, using the inferred $K$,
the natural frequency distribution $g(\omega)$
is reconstructed from the self-consistent susceptibility equations.
This second step reduces to a linear integral equation
of convolution type, which can be efficiently solved
in Fourier space, where the convolution becomes a multiplication.

\subsection{Inference of $K$}
We use the maximum deviation from the limit cycle represented in Eq.~\eqref{eq:r1-max}.
Since $\chi_{\rm c},\chi_{\rm s}$, and $r_{1}^{\rm max}$ are observables,
the coupling constant $K$ is inferred by solving the quadratic equation for $K$,
\begin{equation}
  \left( \dfrac{2\alpha \, r_{1}^{\rm max}}{h} \right)^{2}
  = [1+K\chi_{\rm c}(\omega_{\rm p})]^{2} + [K\chi_{\rm s}(\omega_{\rm p})]^{2}.
\end{equation}
Choosing the solution of $K>0$ from the assumption (iii), we have
\begin{equation}
  K = \dfrac{-\chi_{\rm c}+\sqrt{\chi_{\rm c}^{2}-(\chi_{\rm c}^{2}+\chi_{\rm s}^{2})[1-(2\alpha r_{1}^{\rm max}/h)^{2}]}}{\chi_{\rm c}^{2}+\chi_{\rm s}^{2}}
  \label{eq:inference-K}
\end{equation}
where $\chi_{\rm c}$ and $\chi_{\rm s}$ are evaluated at $\omega_{\rm ex}=\omega_{\rm p}$.


\subsection{Inference of $g(\omega)$}
We define $\chi_{\rm c}^{(0)}$ and $\chi_{\rm s}^{(0)}$ by
\begin{equation}
  \chi_{\rm c}^{(0)} = \dfrac{\pi}{2} g + J_{0},
  \quad
  \chi_{\rm s}^{(0)} = \dfrac{\pi}{2} H[g] + J_{1},
  \label{eq:chic0-chis0}
\end{equation}
which are the susceptibilities for $K=0$ and contain only the natural frequency distribution $g(\omega)$ as the unknown quantity.
The self-consistent equation \eqref{eq:SC-chic-chis} is rewritten into
\begin{equation}
  \begin{pmatrix}
    \chi_{\rm c}^{(0)} \\
    \chi_{\rm s}^{(0)} \\
  \end{pmatrix}
  = \dfrac{1}{(1+K\chi_{\rm c})^{2}+(K\chi_{\rm s})^{2}}
  \begin{pmatrix}
    \chi_{\rm c} + K(\chi_{\rm c}^{2}+\chi_{\rm s}^{2}) \\
    \chi_{\rm s} \\
  \end{pmatrix},
\end{equation}
where the right-hand side consists of observed or inferred quantities.
Therefore, the natural frequency distribution $g(\omega)$ can be inferred
by solving one of the two rows. For instance, the first row is read as
\begin{equation}
  \dfrac{\pi}{2} g(\omega_{\rm ex})
  + \int_{-\infty}^{\infty}
    \dfrac{g(\omega)}{4\alpha^{2}+(\omega-\omega_{\rm ex})^{2}} \, d\omega
  = \varphi_{\rm c}(\omega_{\rm ex}).
  \label{eq:inference-g}
\end{equation}
where
\begin{equation}
  \varphi_{\rm c}(\omega_{\rm ex})
  = \dfrac{\chi_{\rm c}+K(\chi_{\rm c}^{2}+\chi_{\rm s}^{2})}{(1+K\chi_{\rm c})^{2}+(K\chi_{\rm s})^{2}}
\end{equation}
is a known function.
The second row must be equivalent to the first one due to the Kramers-Kronig relation.

The linear integral equation \eqref{eq:inference-g} can be rewritten in the convolution form as
\begin{equation}
  \frac{\pi}{2}\, g(\omega_{\mathrm{ex}}) + \bigl(F * g\bigr)(\omega_{\mathrm{ex}})
  = \varphi_{\mathrm{c}}(\omega_{\mathrm{ex}}),
  \label{eq:conv_form}
\end{equation}
where
\begin{equation}
  F(\omega) = \frac{1}{4\alpha^{2} + \omega^{2}}
  \label{eq:Fomega}
\end{equation}
and $F\ast g$ represents the convolution of $F$ and $g$.
The Fourier transform converts a convolution into a product, and we have
\begin{equation}
  \left[ \dfrac{\pi}{2} + \widetilde{F}(\tau) \right] \widetilde{g}(\tau)
  = \widetilde{\varphi}_{\rm c}(\tau),
  \label{eq:Fourier_conv_form}
\end{equation}
where, for an arbitrary function $\psi(\omega)$, the Fourier transform is defined by
\begin{equation}
  \widetilde{\psi}(\tau)
  = \frac{1}{2\pi} \int_{-\infty}^{\infty}
  \psi(\omega)\, e^{-i\omega\tau}\, d\omega.
  \label{eq:g_fourier}
\end{equation}
For instance, the Fourier transform of $F(\omega)$ is
\begin{equation}
  \widetilde{F}(\tau)
  = \frac{1}{4\alpha}\, e^{-2\alpha|\tau|}.
  \label{eq:Ftau}
\end{equation}
The algebraic equation \eqref{eq:Fourier_conv_form} is easily solved
and $g(\omega)$ is obtained by the inverse Fourier transform of $\widetilde{g}(\tau)$ as
\begin{equation}
  g(\omega)
  = \int_{-\infty}^{\infty}
  \frac{\widetilde{\varphi}_{\mathrm{c}}(\tau)}{\pi/2 + \widetilde{F}(\tau)}
    e^{i\omega \tau}\, d\tau.
  \label{eq:g_omega_inverse}
\end{equation}
When the data in \( \omega_{\mathrm{ex}} \) are discrete, one may implement the same procedure with the discrete Fourier transform; in practice, the fast Fourier transform reduces the computational cost of the Fourier transform $\widetilde{\varphi}_{\rm c}(\tau)$ and of the inverse Fourier transform in \eqref{eq:g_omega_inverse}
for a large set of data, although we will compute in Sec.~\ref{sec:infer-gomega} the transforms directly for a small set of data.

\section{Numerical tests}
\label{sec:NumericalTests}
We set the natural frequency distribution as a normal distribution  
\begin{equation}
  g(\omega)
  = \frac{1}{\sqrt{2\pi \sigma^2}} 
  e^{-(\omega-\mu)^{2}/(2\sigma^{2})},
\label{eq:g_omega}
\end{equation}
where \( \mu = 5 \) and \( \sigma = 0.3 \).
This distribution gives the critical coupling strength
$K_{\rm c}=2/[\pi g(\mu)]=2\sigma\sqrt{2/\pi} \simeq 0.47873$
for the reduced coupled phase-oscillator system,
and $K$ is chosen from the interval $K\in (0,K_{\rm c})$.
The number of oscillators is $N=10^{5}$.
We set $\alpha=1$, which determines
the linear growth ratio from the origin and
the relaxation rate of the limit-cycle amplitude.
Strength of the external force is $h=0$ or $h=0.05$
so as to satisfy the restriction $h\gg 1/\sqrt{N}\simeq 0.00316$.
Frequencies of the external force are sampled from the interval $\omega_{\rm ex}\in [3,7]$
using a discretization step $\Delta\omega_{\rm ex}=0.01$.
The external force is applied from $t=0$. 

The coupled SL oscillator system \eqref{eq:SLmodel}
is integrated using the fourth-order Runge-Kutta method with a time step $dt=0.01$.
The time window \( [t_1, t_2] = [150, 300]\) is used to estimate the susceptibilities
$\chi_{\rm c}$ and $\chi_{\rm s}$ in Eq.~\eqref{eq:chic-chis}
unless otherwise specified.

\subsection{Direct problem}

\begin{figure}[!htbp]
  \centering
  \includegraphics[width=8.5cm]{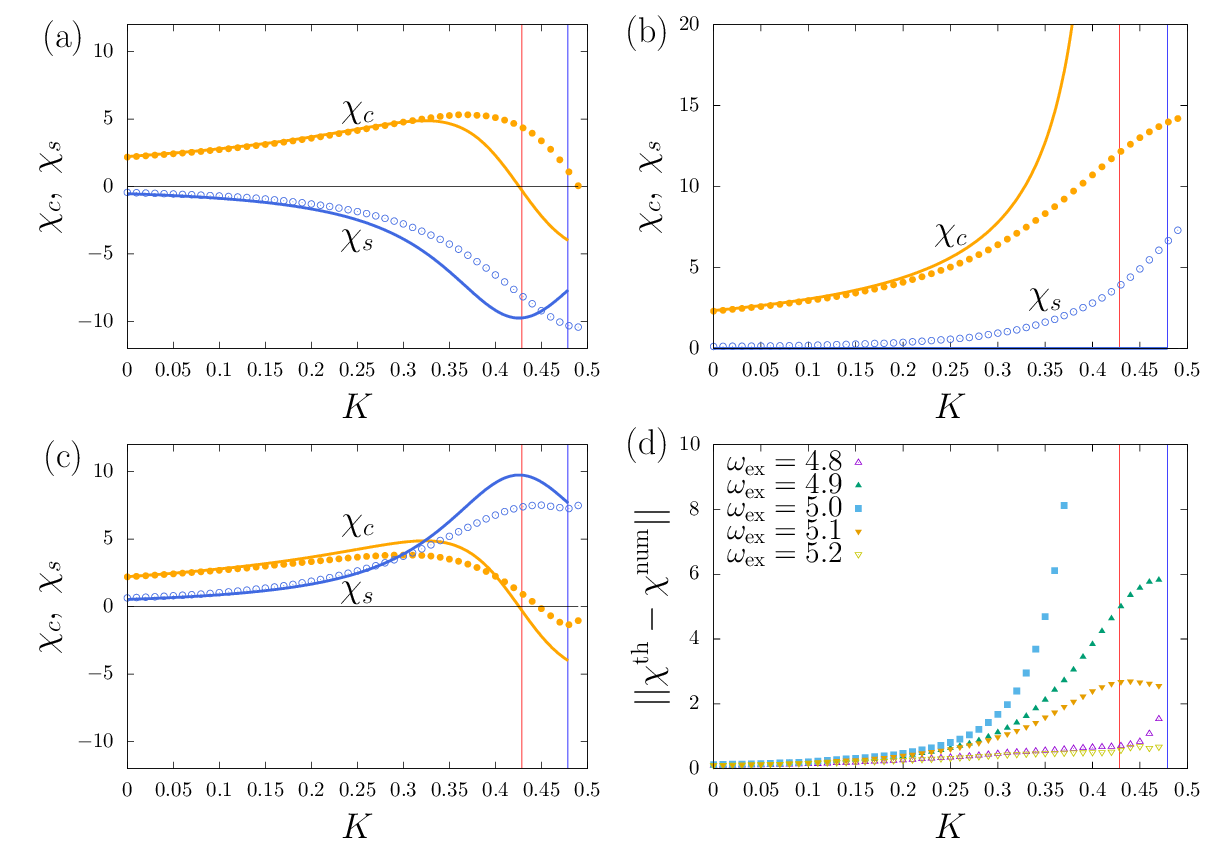}
  \caption{Susceptibilities $\chi_{\rm c}(\omega_\text{ex})$ and $\chi_{\rm s}(\omega_\text{ex})$
    as functions of the coupling constant $K$.
    (a) $\omega_{\rm ex}=4.9$. (b) $\omega_{\rm ex}=5.0$. (c) $\omega_{\rm ex}=5.1$.
    Here $\mu=5.0$ is the symmetry point in $g(\omega)$.
    Theoretical lines for $\chi_{\rm c}$ (orange solid) and $\chi_{\rm s}$ (blue solid).
    Numerical points for $\chi_{\rm c}$ (orange filled circles)
    and $\chi_{\rm s}$ (blue open circles).
    (d) Difference of susceptibility between the theoretical one $\chi^{\rm th}$
    and numerical one $\chi^{\rm num}$.
    Each susceptibility is a two-dimensional vector $\chi=(\chi_{\rm c},\chi_{\rm s})$.
    $\omega_{\rm ex}=4.8$ (purple open triangles),
    $4.9$ (green filled triangles),
    $5.0$ (blue filled squares),
    $5.1$ (orange inverse filled triangles),
    and $5.2$ (yellow inverse open triangles).
    $N=10^5, ~dt=0.01, ~h=0.05$.
    Points are computed from times series in $t \in [100,200]$.
    The blue and red vertical lines mark respectively
    $K_{\rm c}=0.47873$ as the critical point of the reduced phase dynamics
    and $K_{\rm d}=0.42853$ as the divergence point of $\chi^{\rm th}(\mu)$.
}
  \label{fig:chi} 
\end{figure}

We start from examining the validity of the response theory
developed in Sec.~\ref{Susceptibilities}.
The susceptibilities $\chi_{\rm c}$ and $\chi_{\rm s}$ are compared in Fig.~\ref{fig:chi}
between the theory given by Eq.~\eqref{eq:SC-chic-chis}
and observation obtained by Eq.~\eqref{eq:chic-chis}.
They are in good agreement for sufficiently small $K$,
but the agreement becomes worse around the critical point,
in particular with $\omega_{\rm ex}=\mu=5$.
The theory is based on the assumptions of the limits $N\to\infty$ and $h\to 0$,
but our experiments are performed for finite $N$ and finite $h$.
The finiteness induces errors in the susceptibilities $\chi_{\rm c}$ and $\chi_{\rm s}$.

The theoretical $\chi_{\rm c}$ diverges at $K=K_{\rm c}\simeq 0.47873$ and $\omega_{\rm ex}=\mu$
in a coupled phase-oscillator system \cite{Terada2020}.
In the considering coupled SL system,
by fixing $\omega_{\rm ex}=\mu$, we have $H[g](\mu)=J_{1}(\mu)=0$ from symmetry of $g(\omega)$,
and
\begin{equation}
  \chi_{\rm c}(\mu) = \dfrac{\chi_{\rm c}^{(0)}(\mu)}{1-K\chi_{\rm c}^{(0)}(\mu)},
  \quad
  \chi_{\rm s}(\mu) = 0
\end{equation}
from Eq.~\eqref{eq:SC-chic-chis}.
Therefore, the divergence occurs at $K_{\rm d}=1/\chi_{\rm c}^{(0)}(\mu)$,
which is estimated as $K_{\rm d}\simeq 0.42853$ for $\sigma=0.3$.
We note that the divergence does not occurs for $\omega_{\rm ex}\neq\mu$ in general,
since the determinant of the linear equation \eqref{eq:SC-chic-chis} for $(\chi_{\rm c},\chi_{\rm s})$,
that is $(1-K\chi_{\rm c}^{(0)})^{2}+(K\chi_{\rm s}^{(0)})^{2}$, does not vanish.

We remark symmetry of $\chi_{\rm c}(\omega_{\rm ex})$ and $\chi_{\rm s}(\omega_{\rm ex})$.
Theoretically, we have symmetry of $\chi_{\rm c}(\mu+\epsilon)=\chi_{\rm c}(\mu-\epsilon)$
and $\chi_{\rm s}(\mu+\epsilon)=-\chi_{\rm s}(\mu-\epsilon)$ for any $\epsilon\in\mathbb{R}$.
However, this symmetry breaks numerically as shown in Figs.~\ref{fig:chi}(a) and (b).
The symmetry breaking is also observed in Fig.~\ref{fig:chi}(d).

\subsection{Inference of coupling constant $K$}
\label{sec:infer-K}

We randomly pick up an oscillator, say the p-th oscillator,
and observe the times series $x_{\rm p}(t)$ to infer the coupling constant $K$.
The method is two fold:
observation of local maxima of $x_{\rm p}(t)$ without external force and with external force.

\begin{figure}[htb]
  \includegraphics[width=8cm]{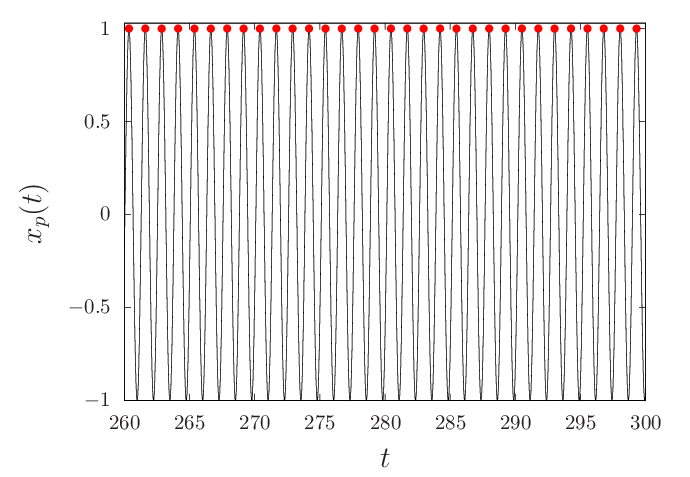}
  \caption{Time series of a sampled oscillator $x_{\rm p}(t)$ (black solid) without external force.
    The red points mark local maxima. $K=0.05$.
    The estimated frequency is $\omega_{\rm p}^{\rm obs} \simeq 5.0015$,
    while the true frequency is $\omega_{\rm p} \simeq 5.0016$.}
  \label{fig:x_p}
\end{figure}

The first step is an observation without external force
to estimate its natural frequency $\omega_{\rm p}$ and the linear growth ratio $\alpha$.
We identify the peaks of $x_{\rm p}(t)$ in the time interval $t\in [t_{1}, t_{2}]$
as shown in Fig.~\ref{fig:x_p}.
The period $T_{\rm p}$ is estimated as $T_{\rm p}^{\rm obs}$ from the average time difference
between two successive peaks, and the frequency $\omega_{\rm p}$ is estimated
by $\omega_{\rm p}^{\rm obs}=2\pi/T_{\rm p}^{\rm obs}$. In addition, the radius of the unperturbed limit cycle $r_{0{\rm p}}$ is estimated from the mean value of the detected peak amplitudes of $x_{\rm p}(t)$. We obtain $r_{0{\rm p}}^{\rm obs} \simeq 0.9999$, which is in close agreement with the theoretical value $r_{0}=1$.
The linear growth ratio $\alpha$ is hence estimated as $\sqrt{\alpha}=0.9999$.

\begin{figure}[!htbp]
  \centering
  \includegraphics[width=8cm]{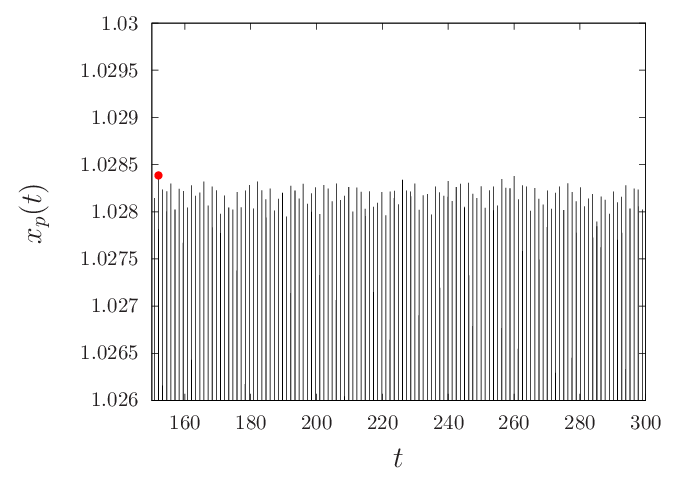}
  \caption{Time series of a sampled oscillator $x_{\rm p}(t)$ (black solid) with external force
    $h=0.05$.
    The red point marks the global maximum.
    $K=0.05$ and $\omega_{\rm ex}\simeq 5.0015$.
  }
  \label{fig:x_p2} 
\end{figure}

In the second step, we apply an external force with the estimated frequency
$\omega_{\rm ex}=\omega_{\rm p}^{\rm obs}$ from $t=0$.
The maximum deviation $r_{1}^{\rm max}$ is evaluated as the difference
between the maximum amplitude of $x_{\rm p}(t)$ in the time interval
$t\in [t_{1}, t_{2}]$ and the unperturbed limit-cycle radius
$r_{0{\rm p}}^{\rm obs}$ obtained in Step~1.
As shown in Fig.~\ref{fig:x_p2} for $K=0.05$ and
$\omega_{\rm ex} \simeq 5.0015$, we obtain
$r_{1}^{\rm max} \simeq 0.0284$.
In another numerical experiment we obtain the susceptibilities from Eq.~\eqref{eq:chic-chis} as
\begin{equation}
  \chi_{\rm c}(\omega_{\rm p}) \simeq 2.6121,
  \quad
  \chi_{\rm s}(\omega_{\rm p}) \simeq 0.1464.
\end{equation}
Substituting these observed values into Eq.~\eqref{eq:inference-K},
we have the inferred value of $K$ as $K\simeq 0.0533$.

\begin{figure}[!htbp]
  \includegraphics[width=8cm]{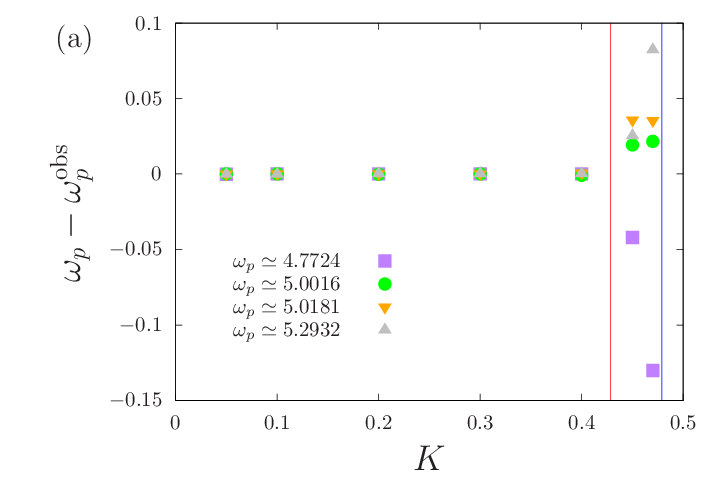}
  \includegraphics[width=8cm]{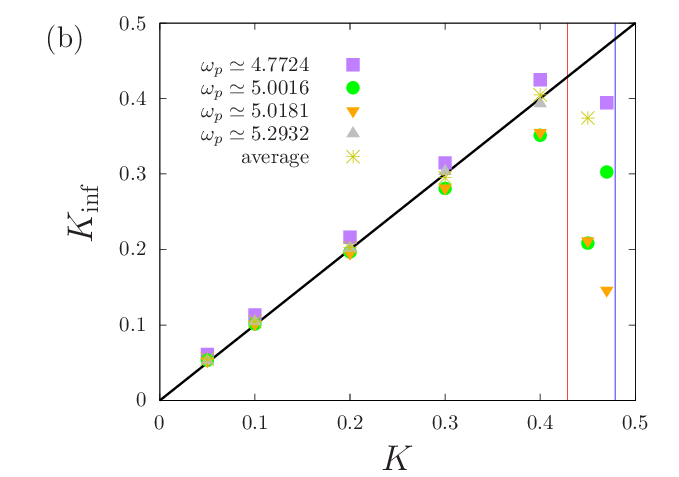}
  \caption{
     Inference of the coupling constant $K$ from 4 oscillators,
     whose frequencies are $\omega_{\rm p} \simeq 4.7724$ (purple squares),
     $5.0016$ (green circles), $5.0181$ (orange inverted-triangles)
     and $5.2932$ (gray triangles). 
    (a) Error of the estimated frequency $\omega_{\rm p}^{\rm obs}$.
    (b) Inferred coupling constant $K_{\rm inf}$.
    The average over the $4$ oscillators is represented by yellow stars.
    The black line represents $K_{\rm inf}=K$.
    The blue and red vertical lines mark respectively
    the critical point $K_{\rm c}\simeq 0.47873$ and the divergence point $K_{\rm d}\simeq 0.42853$.
    The $y$-axis range is restricted to $0 \le K_{\rm inf} \le 0.5$,
    so several outliers are not shown.
    The omitted values are
    for $K>K_{\rm d}$: $(K,K_{\rm inf},\omega_{\rm p})
      = (0.45, 0.6832, 4.7724)$, $(0.45, 1.0838, 5.2932)$, $(0.47, 1.2207, 5.2932)$,
      $(0.47, 1.2207, 5.2932)$, and the average value $K_{\rm inf}=0.5160$ at $K=0.47$.}

\label{fig:inference_K}
\end{figure}

  To enrich statistics, we repeat the above inference of $K$ for other three oscillators
  with varying the true value of $K$.
  First, we examine precision of $\omega_{\rm p}^{\rm obs}$ [see Fig.~\ref{fig:inference_K}(a)].
  Errors are sufficiently small in the interval of $K\in (0,K_{\rm d})$.
  The inferred values of $K$ are reported in Fig.~\ref{fig:inference_K}(b).
  The inference performs well in general,
  although slight errors depending on frequency appear. 

\begin{figure}[!htbp]
  \includegraphics[width=8cm]{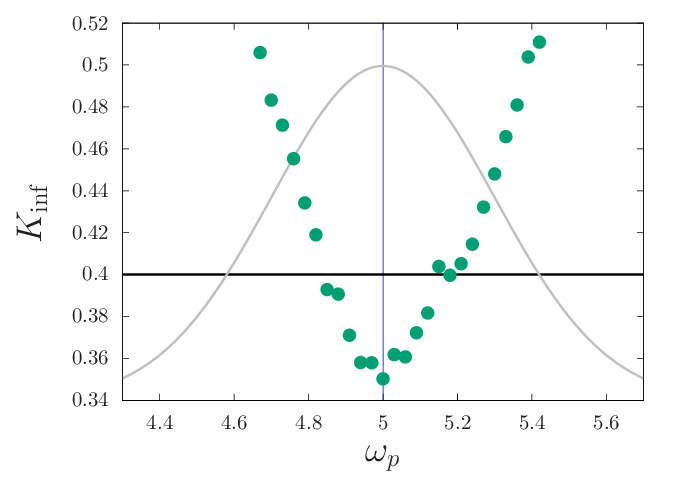}
  \caption{
    Inferred value $K_{\rm inf}$ (green circles)
    with varying the frequency $\omega_{\rm p}$ of a sampled oscillator.
    The black horizontal line marks the true value $K=0.4$.
    The gray curve represents $0.12 g(\omega)+0.34$
    to adjust the scale of panel.
    The blue vertical line marks the symmetry point in $g(\omega)$.
    }
\label{fig:omegap_dep}
\end{figure}

 We investigate $\omega_{\rm p}$ dependence of precision in detail.
 To exclude the error $\omega_{\rm p}-\omega_{\rm p}^{\rm obs}$,
  we use the true value of $\omega_{\rm p}$ and repeat inference with varying $\omega_{\rm p}$.
  The inferred values of $K$ for $K=0.4$ are reported in Fig.~\ref{fig:omegap_dep}.
  We have three remarks.
    First, an oscillator is not suitable to infer $K$ precisely,
    if its frequency $\omega_{\rm p}$ is close to the symmetric point $\mu$ of $g(\omega)$.
  As observed in Fig.~\ref{fig:chi},
  $\omega_{\rm ex}=\mu$ induces divergence of theoretical susceptibility $\chi_{\rm c}$,
  and the divergence induces a larger discrepancy from the observed finite value.
    Second, however, $\omega_{\rm p}$ far from $\mu$ is not suitable neither,
    since the susceptibility becomes small;
    The smallness enhances an error of $K$,
    since $K$ is of order $O(1/|\chi|)$ in the inference formula \eqref{eq:inference-K}.
%
%
Third, as a result, a better agreement between $K$ and $K_{\rm inf}$ is observed
  around $\omega_{\rm ex}\simeq 4.8$ and $5.2$.

\subsection{Natural frequency distribution $g(\omega)$}
\label{sec:infer-gomega}
We now infer the frequency distribution $g(\omega)$ from the measured
response $\varphi_{\rm c}(\omega_{\rm ex})$ where $\omega_{\rm ex}$ is sampled on a uniform grid as $\omega_{\rm ex}^{(1)} = 3, ~\omega_{\rm ex}^{(S)} = 7, ~S = 401$ with the spacing $\Delta\omega_{\rm ex} = 0.01$. For the Fourier transform and its inverse in Eqs. (\ref{eq:g_fourier}) and (\ref{eq:g_omega_inverse}), we discretize them as
\begin{equation}
  \begin{split}
    & \widetilde{\varphi}_{\rm c}(\tau_{k})
      \approx
      \dfrac{\Delta\omega}{2\pi} \sum_{n=1}^{S} \varphi_{\rm c}(\omega_{n}) e^{-i\omega_{n} \tau_{k}}, \\
    & g(\omega_n)
      \approx
      \Delta\tau \sum_{k=1}^{S}\widetilde{g}(\tau_k)e^{i\omega_n\tau_k},
\end{split}
\end{equation}
  where $\omega_{n}=\omega_{\rm ex}^{n}$
  and $\tau_{k}=k\Delta\tau$ with the dual spacing satisfying $\Delta\omega\Delta\tau=2\pi/S$.\\
The imaginary part of the inferred $g(\omega)$ is zero theoretically and its small imaginary part arising from numerical round-off is discarded.

\begin{figure}[!htbp]
  \includegraphics[width=8cm]{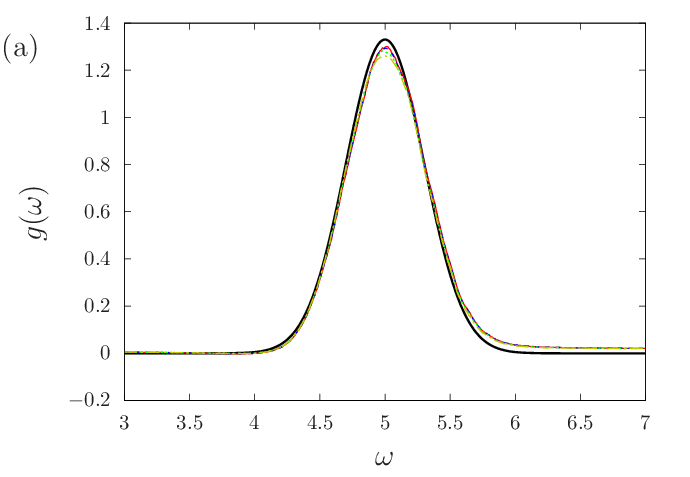}\\
  \includegraphics[width=8cm]{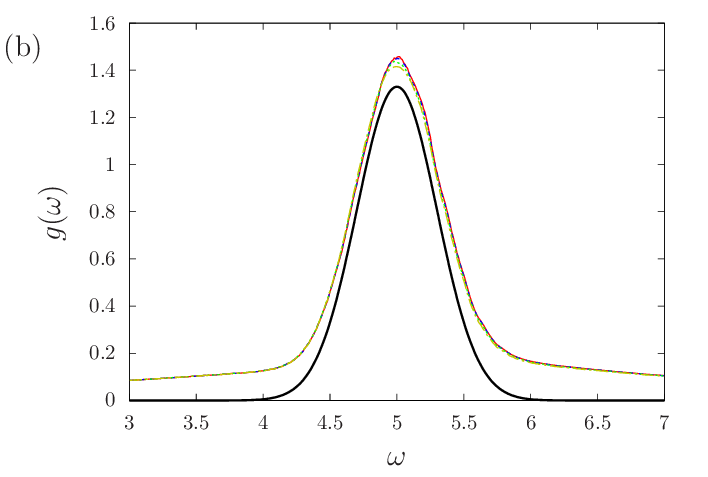}
  \caption{
    Inferred frequency distributions \(g(\omega)\)
    using the average of the inferred coupling strength \(K\)
    reported in Fig.~\ref{fig:inference_K}(b).
    (a) Inference from the total response $\bar{x}=\bar{x}_{\theta}+\bar{x}_{r}$.
    (b) Inference from only the phase response $\bar{x}_{\theta}$.
    In each panel, $K=0.05$ (red solid), $K=0.1$ (blue dashed line), $K=0.2$ (green dotted line), and $K=0.3$ (yellow dash-dotted line) are compared with the ground truth (bold black solid).}
  \label{fig:g_suitei}
\end{figure}

For $K=0.05, 0.1, 0.2$, and 0.3, the inferred natural frequency distribution $g(\omega)$ is shown in Fig.~\ref{fig:g_suitei}. The inference is successful as shown in the panel (a), where the total response $\bar{x}=\bar{x}_{\theta}+\bar{x}_{r}$ is used as discussed in Sec.~\ref{sec:InverseProblemInference}.  In contrast, in the panel (b), the inference is performed by redefining $\chi_{\rm c}$ and $\chi_{\rm s}$ by using $\bar{x}_{\theta}$ instead of $\bar{x}$ in Eq.~\eqref{eq:xt-chi}, in other words, by neglecting the second term in the left-hand side of Eq.~\eqref{eq:inference-g}. This unsuccess implies that it is obligable to consider the amplitude response for a precise inference in coupled limit-cycle systems. We note that, even in the panel (a), the peak height of the inferred distribution is slightly lower than the true one. This deviation is mainly attributed to a small estimation error in the coupling strength $K$, which propagates into the inference of $g(\omega)$.

\section{Conclusion}
\label{sec:Conclusion}
In this study, we proposed a methodology for inferring system parameters in Stuart-Landau (SL) oscillator systems subjected to weak external forcing. The central idea is to utilize the linear response of a macroscopic variable to external periodic input, combined with microscopic observations from a small number of oscillators.

The microscopic observations infer the coupling constant $K$ with the aid of observed macroscopic responses.
Using the inferred value of $K$, the natural frequency distribution $g(\omega)$ is inferred
from the macroscopic responses by varying the external frequency $\omega_{\rm ex}$.
A main computing cost comes from the inference of $g(\omega)$.
However, the cost can be drastically reduced due to formulation in the Fourier transform
and utilization of the fast Fourier transform,
even if the number of observation points $\omega_{\rm ex}$ is large.
Overall, the proposed approach provides a practical and accurate method for inferring both the natural frequency distribution and coupling strength from partial observations in SL systems.
The methodology offers a tractable framework for inverse problems in coupled oscillator systems.

  From remarks made in Sec.~\ref{sec:infer-K}, to infer the coupling constant $K$,
  it is advisable to exclude oscillators with natural frequencies
  located around the center or in tails of the natural frequency distribution $g(\omega)$.
  Actually, we have to know $g(\omega)$ to exclude such oscillators,
  while $K$ is necessary to infer $g(\omega)$.
  We need to develop a self-consistent manner to improve precision of the inferences.


Beyond the Stuart-Landau model, many other limit-cycle systems, such as the van der Pol oscillator or the FitzHugh-Nagumo model, present additional challenges.
The phase reduction can be systematically carried out,
and the linear response theory shown in Appendix \ref{sec:LinearResponsePhase} can be
applied to general phase-oscillator systems.
A remaining part is the response in amplitude,
since the SL oscillator simplifies the solution obtained
in Appendix \ref{sec:LinearResponseAmplitude},
although a systematic extension may be possible as discussed in Appendix \ref{sec:PhaseAmplitudeEqs}.
Extending the current inference framework to such systems will require further theoretical
developments and remains an important direction for future researches.

\acknowledgements
 MC was supported by JST SPRING, Japan Grant Number JPMJSP2180. YYY acknowledges the support of JSPS KAKENHI Grant No. JP21K03402.

\appendix

\section{Linear response of phases}
\label{sec:LinearResponsePhase}

We consider the equation
\begin{equation}
  \dot{\theta}_{j} = \omega_{j} - \frac{h}{\sqrt{\alpha}} H(\theta_{j},t)
  \qquad
  (j=1,\cdots,N)
\end{equation}
where
\begin{equation}
  H(\theta,t) = A\sin(\theta-\omega_{\rm ex}t) + B\cos(\theta-\omega_{\rm ex}t).
\end{equation}
In the limit $N\to\infty$, the equation of continuity is
\begin{equation}
  \dfracp{F}{t} + \dfracp{}{\theta}(VF) = 0,
\end{equation}
and the velocity $V$ is
\begin{equation}
  V(\theta,\omega,t) = \omega - \frac{h}{\sqrt{\alpha}} H(\theta,t).
\end{equation}
Here, $F(\theta,\omega,t)d\theta d\omega$ is the fraction of oscillators
which are in the region $[\theta,\theta+d\theta]\times [\omega,\omega+d\omega]$ at time $t$.
We compute the linear response to the external force 
{$-hH/\sqrt{\alpha}$
around the nonsynchronized stationary state $F_{0}(\omega)=g(\omega)/(2\pi)$.

Expanding $F=F_{0}+hf$, the linearized equation of continuity becomes
\begin{equation}
  \dfracp{f}{t} + \omega \dfracp{f}{\theta} - \frac{1}{\sqrt{\alpha}} F_{0} \dfracp{H}{\theta} = 0.
\end{equation}
We introduce the Fourier-Laplace transform of $f$ as
\begin{equation}
  \widehat{f}_{n}(\omega,s)
  = \int_{0}^{\infty} dt~ e^{-st} \dfrac{1}{2\pi} \int_{0}^{2\pi} d\theta~
  e^{-in\theta} f(\theta,\omega,t),
\end{equation}
where ${\rm Re}(s)>0$ to ensure convergence of the Laplace transform.
We define the Fourier-Laplace transform $\widehat{H}_{n}$ of $H$ similarly.
The linearized equation is transformed to
\begin{equation}
  (s+in\omega) \widehat{f}_{n}(\omega,s)
  = \frac{in}{\sqrt{\alpha}} F_{0}(\omega) \widehat{H}_{n}(s),
\end{equation}
where we assumed that $f(\theta,\omega,t=0)=0$. We have
\begin{equation}
  \widehat{f}_{n}(\omega,s) = \frac{inF_{0}(\omega)}{\sqrt{\alpha}(s+in\omega)} \widehat{H}_{n}(s).
\end{equation}

We are interested in the order parameter
\begin{equation}
  z(t) = \int_{0}^{2\pi} d\theta \int_{-\infty}^{\infty} d\omega ~e^{i\theta} f(\theta,\omega,t),
\end{equation}
whose Laplace transform is
\begin{equation}
  \widehat{z}(s) = 2\pi \int_{-\infty}^{\infty} \widehat{f}_{-1}(\omega,s) d\omega
  = \frac{1}{\sqrt{\alpha}} \int_{-\infty}^{\infty} \dfrac{-ig(\omega)}{s-i\omega} d\omega~ \widehat{H}_{-1}(s),
\end{equation}
where
\begin{equation}
  \hat{H}_{-1}(s) = \dfrac{iA+B}{2(s-i\omega_{\rm ex})}.
\end{equation}
Temporal evolution $z(t)$ is obtained through the inverse Laplace transform
\begin{equation}
  z(t) = \int_{\rm Br} \dfrac{ds}{2\pi i} e^{st} \hat{z}(s)
\end{equation}
where ${\rm Br}$ represents the Bromwich contour, which runs from $\sigma-i\infty$ to $\sigma+i\infty$
and $\sigma>0$ is taken so that all the singularities of $\widehat{z}(s)$ are in the left.
By adding a half circle in the left side of the Bromwich contour,
which gives zero contribution, the residue theorem picks up the poles.
Picking up the pole at $s=i\omega_{\rm ex}+0$ from the external force, where $+0$ comes from the condition ${\rm Re}(s)>0$, we have 
\begin{equation}
  z(t) =
  \frac{iA+B}{2\sqrt{\alpha}} e^{i\omega_{\rm ex}t}
  \int_{-\infty}^{\infty} \dfrac{g(\omega)}{\omega-\omega_{\rm ex}-i0} d\omega.
\end{equation}
The integral is transformed into
\begin{equation}
  \int_{-\infty}^{\infty} \dfrac{g(\omega)}{\omega-\omega_{\rm ex}-i0} d\omega
  = - \pi H[g](\omega_{\rm ex}) - i\pi g(\omega_{\rm ex}).
\end{equation}
The final form of $z(t)$ is
\begin{equation}
  z(t) =
  \frac{(A-iB)\pi}{2\sqrt{\alpha}} \left\{ g(\omega_{\rm ex}) - i H[g](\omega_{\rm ex}) \right\}
  e^{i\omega_{\rm ex}t}.
\end{equation}
The macroscopic variable $\bar{x}_{\theta}$ is the real part of $z(t)$
multiplied by $\sqrt{\alpha}$, and we have
\begin{equation}
  \bar{x}_{\theta}(t)
  = h \dfrac{\pi}{2} \left[
    ( Ag-BH[g] ) \cos\omega_{\rm ex}t
    + (AH[g]+Bg ) \sin\omega_{\rm ex}t
  \right].
\end{equation}

\section{Linear response of amplitudes}
\label{sec:LinearResponseAmplitude}
We solve Eq.~\eqref{eq:EOM-r-2} with approximating $\theta_{j}\simeq\omega_{j}t+\delta_{j}$,
where $\delta_{j}$ is the initial phase of $\theta_{j}$.
The approximated solution is $r_{1j}(t)\xrightarrow{t\to\infty}r_{1}(t;\omega_{j},\delta_{j})$,
where
\begin{equation}
  r_{1}(t;\omega,\delta)
  = h \left[
    \dfrac{2\alpha A+B\Delta\omega}{4\alpha^{2}+\Delta\omega^{2}} C(t)
    + \dfrac{A\Delta\omega-2\alpha B}{4\alpha^{2}+\Delta\omega^{2}} S(t)
  \right].
  \label{eq:r1-t-omega-delta}
\end{equation}
$\Delta\omega=\omega-\omega_{\rm ex}$, and
\begin{equation}
  C(t) = \cos(\Delta\omega t+\delta),
  \quad
  S(t) = \sin(\Delta\omega t+\delta).
\end{equation}

Recall the definition \eqref{eq:definition-xtheta-xr} of the response of amplitudes $\bar{x}_{r}$.
Replacing the $N$-body average with the average over the nonsynchronized
initial distribution $F_{0}(\omega)=g(\omega)/2\pi$,
the macroscopic response $\bar{x}_{r}$ is obtained as
\begin{equation}
  \bar{x}_{r}
  \xrightarrow{t\to\infty}
  \int_{-\infty}^{\infty} d\omega \int_{0}^{2\pi} d\delta~
  \dfrac{g(\omega)}{2\pi} ~ r_{1}(t;\omega,\delta) \cos(\omega t+\delta)
  + O(h^{2}),
\end{equation}
where we approximated $\theta(t)\simeq\omega t+\delta$.
Performing the average over $\delta$, we have Eq.~\eqref{eq:barx-r}.

From Eq.~\eqref{eq:r1-t-omega-delta}, 
the maximum value of an individual oscillator is
\begin{equation}
  \begin{split}
    r_{1}^{\rm max} 
    & = h \sqrt{ \left( 
      \dfrac{2\alpha A+B\Delta\omega}
            {4\alpha^{2}+\Delta\omega^{2}} 
      \right)^{2}
      + \left(
      \dfrac{A\Delta\omega-2\alpha B}
            {4\alpha^{2}+\Delta\omega^{2}}
      \right)^{2} } \\
    & = h \sqrt{\dfrac{A^{2}+B^{2}}
                     {4\alpha^{2}+\Delta\omega^{2}} }.
  \end{split}
\end{equation}
In particular, setting $\omega_{\rm ex}=\omega$, the maximum value is simplified to
\begin{equation}
  r_{1}^{\rm max}
  = \dfrac{h}{2\alpha}
    \sqrt{A^{2}+B^{2}}.
  \label{eq:r-max}
\end{equation}

\section{Phase-amplitude reduction}
\label{sec:PhaseAmplitudeEqs}

Let us consider a perturbed oscillator described by
\begin{equation}
  \dot{\boldsymbol{X}} = \boldsymbol{F}(\boldsymbol{X}) + \epsilon\boldsymbol{p}(t),
\end{equation}
where $\boldsymbol{X}\in\mathbb{R}^{n}$ is the state,
$\boldsymbol{F}(\boldsymbol{X})$ represents the unperturbed system
having a limit-cycle $\boldsymbol{U}(t)$,
and $\epsilon\boldsymbol{p}(t)$ is perturbation.
The perturbation may include coupling terms and external forces.
We have a coordinate transform from $\boldsymbol{X}$ to
$\boldsymbol{Y}=(\theta,\rho_{2},\cdots,\rho_{n})$,
where $\theta$ is the phase defined by the isochrons,
and $\rho_{i}$ are the amplitudes defined by the isostables
\cite{Mauroy2013,Mauroy2014,Wilson2015,Wilson2016}.

To implement the phase-amplitude reduction,
we consider the time-$t$ Koopman operator, which is linear and denoted by $\mathcal{K}^{t}$.
We assume that $\mathcal{K}^{t}$ has eigenvalues $e^{\lambda_{1}t},\cdots,e^{\lambda_{n}t}$
and associated eigenfunctions $s_{1}(\boldsymbol{X}),\cdots,s_{n}(\boldsymbol{X})$.
The phase-amplitude equations around the limit cycle are read as \cite{Shirasaka2017}
\begin{equation}
  \begin{split}
    & \dot{\theta} = \omega + \epsilon \nabla\Theta(\boldsymbol{U}(t)) \cdot
      \boldsymbol{p}(t), \\
    & \dot{\rho}_{i} = \lambda_{i} \rho_{i} + \epsilon \nabla R_{i}(\boldsymbol{U}(t)) \cdot \boldsymbol{p}(t),
      \qquad (i=2,\cdots,n) \\
  \end{split}
  \label{eq:PAEgeneral}
\end{equation}
where $\lambda_{1}=i\omega$,
$\Theta(\boldsymbol{X})={\rm arg}(s_{1}(\boldsymbol{X}))$,
and $R_{i}(\boldsymbol{X})={\rm Re}(s_{i}(\boldsymbol{X}))$.
The phase-amplitude equations are linear inhomogeneous equations
as a Stuart-Landau system analyzed in this article.
Our job is hence to compute the eigenvalues and eigenfunctions of the Koopman operator.
We may search eigenvalues $\lambda_{1},\cdots,\lambda_{n}$
  and associated eigenfunctions $s_{1}(\boldsymbol{X}),\cdots,s_{n}(\boldsymbol{X})$
  of the Lie operator expressed by
\begin{equation}
  \mathcal{L} = \boldsymbol{F}(\boldsymbol{X})\cdot\nabla,
\end{equation}
instead of the Koopman operator $\mathcal{K}^{t}$.

Let the system $\boldsymbol{F}(\boldsymbol{X})$
be the Stuart-Landau system \eqref{eq:SLmodel-mono} with $\alpha=1$,
which is read as
\begin{equation}
  \boldsymbol{F}(\boldsymbol{X}) = 
  \begin{pmatrix}
    x - \omega y - (x^{2}+y^{2})x \\
    y + \omega x - (x^{2}+y^{2})y \\
  \end{pmatrix},
\end{equation}
and confirm that the general framework \eqref{eq:PAEgeneral}
recovers our phase-amplitude equations obtained through the polar coordinates
\begin{equation}
  \begin{split}
    & \dot{\theta} = \omega - \epsilon p(t) \sin\theta, \\
    & \dot{r}_{1} = -2r_{1} + \epsilon p(t) \cos\theta, \\
  \end{split}
  \label{eq:PAE}
\end{equation}
where $r_{1}$ is the deviation from the limit cycle.
See Eqs.~\eqref{eq:EOM-theta} and \eqref{eq:EOM-r1}.
We can take eigenfunctions $s_{1}(x,y)$ and $s_{2}(x,y)$ as
\begin{equation}
  s_{1}(x,y) = \dfrac{x+iy}{r},
  \quad
  s_{2}(x,y) = \dfrac{\sqrt{(1-r^{2})^{2}}}{2r^{2}} =: S_{2}(r)
\end{equation}
with $r=\sqrt{x^{2}+y^{2}}$, where $\mathcal{L}s_{1}=i\omega s_{1}$
and $\mathcal{L}s_{2}=-2 s_{2}$.
They define $\Theta(x,y)$ and $R_{2}(x,y)$, and their gradients are
\begin{equation}
  \nabla\Theta(\boldsymbol{X}) =
  \begin{pmatrix}
    -\sin\theta \\ \cos\theta
  \end{pmatrix},
  \quad
  \nabla R_{2}(\boldsymbol{X})
  = \dfrac{{\rm sgn}(r^{2}-1)}{r^{3}}
  \begin{pmatrix}
    \cos\theta \\ \sin\theta
  \end{pmatrix},
\end{equation}
where $(x,y)=(r\cos\theta,r\sin\theta)$.
Since the limit cycle is $r=1$ and the amplitude $\rho_{2}$ relates to the deviation $r_{1}$ as
\begin{equation}
  \rho_{2} = S_{2}(1+r_{1}) = |r_{1}| + O(r_{1}^{2}),
\end{equation}
we recover the phase-amplitude equations \eqref{eq:PAE}
in our setting $\boldsymbol{p}(t)=(p(t),0)$.

\end{document}